\documentstyle[preprint,aps]{revtex}
\begin{document}
\draft
\title{ Vacuum engineering with DCC}

\author{Ji\v{r}\'{\i} Ho\v{s}ek \\
Nuclear Physics Institute, Czech Acad.Sci., \\
250 68 \v{R}e\v{z} (Prague), Czech Republic }
\maketitle

\begin{abstract}
Formation of a vacuum like domain of the strong-interaction chiral condensate
misaligned with respect to the electroweak symmetry breaking arguably modifies
the mass spectrum of electroweak gauge bosons and their interactions in that
domain in an observable manner. In particular, photon acquires the mass
$m_{\gamma}\sim \frac{1}{2} f_{\pi}\sin 2 \theta_{W}$.

\end{abstract}
\newpage

General idea of the vacuum engineering [1] finds its explicit
realization in the concept of the disordered chiral condensate (DCC) suggested
to form in the high-energy hadronic or nuclear collisions [2]. The experimental
signals discussed in the literature for its manifestation are very interesting:
(1) Perhaps the most robust would be the observation of the anomalous isospin
distribution of pions in the DCC decay [3]

\begin{equation}
P(r) = 1/2 \sqrt{r}
\end{equation}
where $r = n_{\pi^{0}}/(n_{\pi^{0}} + n_{\pi^{+}} + n_{\pi^{-}})$. (2)
There should also be observable effects in processes involving both strong and electromagnetic
interactions. In particular, the dilepton decays of the hadronic resonances
in DCC would exhibit distinguished properties [4].

In this paper we analyze the influence of the misaligned vacuum-like DCC
on the electroweak gauge-boson mass spectrum. In the standard world the effect
of the chirally noninvariant QCD vacuum upon the W and Z boson masses is
known [5] and almost entirely ignored as unobservable [6]. We believe that the
Nambu-Goldstone (NG) pions of DCC "skrewed" with respect to the electroweak
symmetry breaking influence the electroweak gauge-boson mass spectrum in DCC
in an observable manner.

In the standard model the electroweak gauge bosons $W^{\pm}$ and Z acquire masses
primarily by absorbing three NG bosons contained in the complex Higgs doublet
which is supposed to condense in the electroweak vacuum with $v{\cong} 246 GeV$:

\[
<O_{ew} |\phi(x)|O_{ew}> = \frac{1}{\sqrt{2}} v
\]
There is, however, yet another contribution to $m_{W}$ and $m_{Z}$. It is due
to pions [5] by virtue of nonzero matrix elements of the electroweak quark
currents between the strong vacuum and three NG massless pions of spontaneously
broken global $SU(2)_{L}\times SU(2)_{R}$ symmetry of strong interactions
$(f_{\pi}\cong 92.4 MeV)$:

\begin{equation}
<O_{s}|\bar{q} \gamma^{\mu}(1-\gamma_{5})\frac{1}{2} \tau_{i}q|\pi_{j}(k)> =
i\delta_{ij}f_{\pi}k^{\mu}
\end{equation}
As a result,

\[
m^{2}_{W} = \frac{1}{4} g^{2}(v^{2} + f^{2}_{\pi})
\]
\[
m^{2}_{Z} = \frac{1}{4} (g^{2} + g'^{2}) (v^{2} + f^{2}_{\pi}).
\]
Absorbed by W and Z to become their longitudinal polarization components are thus
the linear combinations of the Higgs NG bosons $\theta_{i}$ and the QCD pions
$(f_{\pi}/\sqrt{v^{2} +
f^{2}_{\pi}}\sim 4.10^{-4})$:

\begin{equation}
|\Theta_{i}> = \frac{v}{\sqrt
{v^{2}+f_{\pi}^{2}}}|\theta_{i}>
 + \frac{f_{\pi}} {\sqrt{ v^{2}
+f_{\pi}^{2}}} |\pi_{i}^{QCD}>
\end{equation}
Physical pions which remain in the spectrum are the states orthogonal to (3):

\begin{equation}
|\pi_{i}> = - \frac{f_{\pi}}{\sqrt {v^{2}+f_{\pi}^{2}}}
|\theta_{i}> +
\frac{v}{\sqrt {v^{2}+f_{\pi}^{2}}}| \pi_{i}^{QCD}>
\end{equation}
They are predominantly the QCD pions with a tiny admixture of the Higgses.
To the best of our knowledge there is as yet no experimentally feasible
possibility to distinguish the two contributions [6].

The concept of DCC offers new possibilities: We perform the general
chiral rotation
\begin{eqnarray}
q_{L}\rightarrow exp [i\vec{\theta}_{L}\vec{\tau}]q_{L}
\nonumber
\\
q_{R}\rightarrow exp [i\vec{\theta}_{R}\vec{\tau}]q_{R}
\end{eqnarray}
of the $SU(2)_{L} \times U(1)_{Y}$ gauge interaction of the u,d
quarks

\[
{\cal L}_{int} = g\bar{q}_{L}\gamma^{\mu}\frac{1}{2} \tau_{i}q_{L} A_{i\mu} +
g'[\frac{1}{3}\bar{q}\gamma^{\mu}q + \bar{q}_{R}\gamma^{\mu}\frac{1}{2}\tau_{3}
q_{R}]B_{\mu}
\]
and take the matrix elements (2) of ${\cal L}_{int}$. They correspond to
a simple effective London Lagrangian

\begin{equation}
{\cal L}_{L} = - \frac{1}{2}gf_{\pi}A_{i\mu}L_{ij}\partial ^{\mu}\pi_{j} +
\frac{1}{2} g'f_{\pi}B_{\mu}R_{i}\partial^{\mu}\pi_{i}
\end{equation}
where the matrix L equals
\[
\left[
\begin{array}{ccc}
c_{L}+\hat{\theta}^{2}_{1L}(1-c_{L}) &
\hat{\theta}_{3L}s_{L}+\hat{\theta}_{1L}\hat{\theta}_{2L}(1-c_{L}) &
-\hat{\theta}_{2L}s_{L}+\hat{\theta}_{1L}\hat{\theta}_{3L}(1-c_{L}) \\
-\hat{\theta}_{3L}s_{L}+\hat{\theta}_{1L}\hat{\theta}_{2L}(1-c_{L}) &
c_{L}+\hat{\theta}^{2}_{2L}(1-c_{L}) &
\hat{\theta}_{1L}s_{L}+\hat{\theta}_{2L}\hat{\theta}_{3L}(1-c_{L}) \\
\hat{\theta}_{2L}s_{L}+\hat{\theta}_{1L}\hat{\theta}_{3L}(1-c_{L}) &
-\hat{\theta}_{1L}s_{L}+\hat{\theta}_{2L}\hat{\theta}_{3L}(1-c_{L}) &
c_{L}+\hat{\theta}^{2}_{3L}(1-c_{L})
\end{array}
\right]
\]

\[
R= [\hat{\theta}_{2R}s_{R}+\hat{\theta}_{1R}\hat{\theta}_{3R}(1-c_{R}),
-\hat{\theta}_{1R}s_{R}+\hat{\theta}_{2R}\hat{\theta}_{3R}(1-c_{R}),
c_{R}+\hat{\theta}^{2}_{3R}(1-c_{R})],
\]
\[
c_{L,R} = cos2\theta_{L,R}, s_{L,R} = sin2\theta_{L,R},
\]
\[
\theta_{L,R} =|\vec{\theta}_{L,R}|, \hat{\theta}_{iL,R} = \theta_{iL,R}/
|\theta_{L,R}|,
\]
\begin{equation}
LL^{T}=1,
R^{2}_{i} = 1.
\end{equation}

The gauge-boson mass matrix equals [7] the minus residue at the simple
massless pole of the transverse gauge-boson polarization tensor

\[
i\Pi^{\mu\nu}_{ab}(k) = - i(k^{\mu}k^{\nu} - k^{2}g^{\mu\nu})\Pi_{ab}(k^{2})
\]
Calculating explicitly its longitudial part using the London Lagrangian (6)
(the $g^{\mu\nu}$ term follows by transversality), we obtain the gauge-
boson mass matrix due to DCC in the basis ($A_{1},A_{2},A_{3},B$) as

\[
m^{2}_{DCC}= \frac{1}{4} f_{\pi}^{2} \left[
\begin{array}{cccc}
 g^{2} & 0 & 0 & -gg'(LR)_{1} \\
0 & g^{2} & 0 & -gg'(LR)_{2} \\
0 & 0 & g^{2} & -gg'(LR)_{3} \\
-gg'(LR)_{1} & -gg'(LR)_{2} & -gg'(LR)_{3} & g'^{2}
\end{array} \right ]
\]
It follows from (7) that one eigenvalue of $m^{2}_{DCC}$ equals zero, and
since $m^{2}_{DCC} = MM^{T}$ it also follows that the other three eigenvalues
of $m^{2}_{DCC}$ are manifestly positive; all as expected.

The full electroweak gauge-boson mass matrix in the same basis has the form
\[
m^{2} = \frac{1}{4}v^{2} \left(
\begin{array} {cccc}
g^{2} & 0 & 0 & 0  \\
0 & g^{2} & 0 & 0 \\
0 & 0 & g^{2} & -gg' \\
0 & 0 & -gg' & g'^{2}
\end{array}
\right )
+m^{2}_{DCC}
\]
which, upon diagonalization by an orthogonal $4 \times 4 $ matrix,gives rise
to four massive gauge boson eigenstates.
Taken literally the physical consequences of the DCC dynamics are rather
visible. (i) Due to the charge properties of the DCC vacuum which we examplify
by performing the rotation (5) of the quark vacuum condensate
$ (U^{+}(\theta_{L})U(\theta_{R})\equiv U(\theta))$

\[
<\bar{q}_{L}q_{R}>\rightarrow <\bar{q}_{L}q_{R}> cos\theta + <\bar{q}_{L}
\vec{\tau}q_{R}>i\hat{\vec{\theta}}sin \theta
\]
it is not surprising that the general gauge boson mass eigenstates are
not the eigenstates of the electric
charge. The "wrong" admixtures are of course small of the order $f_{\pi}/
\sqrt{v^{2}+f_{\pi}^{2}}$ and the light mass must be of the order $O(gf_{\pi})$.
(ii) The electroweak interactions say, of leptons, in DCC yield specific
phenomena. Perhaps most dramatic are the parity-violating long-range force, and
the electric charge nonconserving decays like ${\mu}\rightarrow{\nu}_{\mu}$+
light gauge boson.
(iii) Counting the degrees of freedom we conclude that there are only two
massless NG pions left in the physical spectrum of the DCC electroweak
world. Clearly, the relation (1) ignores the electroweak interactions.

We point out that a very similar picture emerges if, besides the standard
electroweak symmetry breaking sector (and the correctly oriented QCD chiral
condensate) new Higgs fields are added which mimic a superconducting domain [8].

For an illustration we specify the general chiral rotation in such a way
that the resulting DCC condensate generates "only" the $Z-\gamma$ mixing.
Also such a possibility was contemplated in the superconducting scenario [9].
Taking first $\theta_{iL} = -\theta_{iR} = -\theta_{i} (\theta_{L} = \theta_{R}
= \theta)$
which gives $(LR)_{k} = [\hat{\theta}_{1}\hat{\theta}_{3}(1-cos2\theta)+
\hat{\theta}_{2}sin2\theta,
\hat{\theta}_{2}\hat{\theta}_{3}(1-cos2\theta)-\hat{\theta}_{1}sin2\theta,
\theta^{2}_{3}(1-cos2\theta)
+cos2\theta]$ we further specify the rotation by taking $\theta_{3}=0,\theta=
\pi/2$. As a result the charged $W^{\pm}$ gauge bosons acquire the mass
$m^{2}_{{W}_{DCC}}=\frac{1}{4}g^{2}(v^{2}+f_{\pi}^{2})=m^{2}_{W}$.
The other two gauge boson mass
eigenstates are obtained by diagonalization of the mass matrix in the $(A_{3},B)$
basis of the form

\[
m^{2} = \frac {1}{4} (v^{2}+f_{\pi}^{2}) \left(
\begin{array} {cc}
g^{2} & -gg'(1- \xi) \\
-gg'(1-\xi) & g'^{2}
\end{array}
\right)
\]
where ${\xi} = 2f^{2}_{\pi}/(v^{2}+f_{\pi}^{2})$.
The mass eigenstates

\[
Z^{\mu}_{DCC} = cos\theta_{DCC}A^{3}_{\mu}+sin\theta_{DCC}B^{\mu}
\]
\[
A^{\mu}_{DCC} = - sin\theta_{DCC}A^{3}_{\mu}+cos\theta_{DCC}B^{\mu}
\]
have their corresponding masses given as
\[
m^{2}_{{Z}_{DCC}} =
\frac{1}{4}(g^{2}+g'^{2})(v^{2}+f_{\pi}^{2})-\frac{g^{2}g'^{2}}
{g^{2}+g'^{2}} f_{\pi}^{2}= m^{2}_{Z} -\frac{1}{4}f^{2}_{\pi}sin^{2}2\theta_{W}
\]
\[
m^{2}_{{A}_{DCC}}= \frac{g^{2}g'^{2}} {g^{2}+g'^{2}} f_{\pi}^{2} =
\frac{1}{4}f_{\pi}^{2} sin^{2}2\theta_{W}
\]
Consequently, in DCC the standard ratio $m^{2}_{W}/m^{2}_{Z}=cos^{2}\theta_{W}$
is modified accordingly.
The mixing angle $\theta_{DCC}$ is related to the Weinberg angle $(tg\theta_{W}=
g'/g)$:

\begin{equation}
tg2\theta_{DCC} = (1-{\xi})tg2\theta_{W}
\end{equation}
It is convenient to express the standard Z and A as mixtures of the corresponding
DCC mass eigenstates:

\begin{eqnarray}
Z^{\mu} = cos (\theta_{DCC}-\theta_{W}) Z^{\mu}_{DCC} -sin(\theta_{DCC}-\theta_{W})
A^{\mu}_{DCC}
\nonumber
\\
A^{\mu} = sin(\theta_{DCC}-\theta_{W})Z^{\mu}_{DCC} + cos(\theta_{DCC}-\theta_{W})
A^{\mu}_{DCC}
\end{eqnarray}
in which the mixings can be expressed in terms of the measurable quantities
$\theta_{W},f_{\pi},m_{Z}$ and e. Not surprisingly the relation (8) implies

\[
\sin(\theta_{DCC}-\theta_{W}) \cong\frac{1}{\sqrt{2}} \frac{f_{\pi}}{m_{Z}} \frac
{e}{sin2\theta_{W}}
\]
The "electromagnetic" interaction of the charged leptons and the
neutral-current interaction of leptons in DCC expressed in terms of
$Z^{\mu}_{DCC}$ and $A^{\mu}_{DCC}$ using (9) yield the total width of the massive
DCC photon approximately
\[
\Gamma\cong\frac{1}{3}{\alpha}m_{{A}_{DCC}}.
\]

The mass $m_{{A}_{DCC}}=f_{\pi}sin{\theta}_{W}cos{\theta}_{W}\cong 38 MeV$
determines the minimal spatial size L of the vacuum-like domain in which the modification
of the electroweak gauge-boson mass spectrum is justified: $ L>5 fm$.

Important question whether the results of a simple calculation presented
here apply to any conceivable experimental situation remains unanswered
at present. Feasibility of the formation of the baked Alaska in high-energy
hadron or heavy-ion collisions which according to our understanding would be
suitable for applying present calculation, is being analyzed both theoretically
[2] and experimentally [10] with unconclusive results. Be it as
it may, we believe that the experiments dedicated to reveal DCC in baked Alaska
are those which at the same time might reveal the QCD contribution
to the electroweak gauge-boson masses.

Present work was done during the author's visit of CERN-TH
Division supported by the ATLAS budget of the Committee for Czech
Republic-CERN Cooperation.

\end{document}